\documentclass{article}
\usepackage{smc2025}
\usepackage[caption=false, font=footnotesize]{subfig}
\usepackage{paralist}
\usepackage[figure,table]{hypcap}

\usepackage[whole]{bxcjkjatype}

\usepackage{alphabeta}
\usepackage{arabtex}
\usepackage[LFE,LAE,LGR,T2A,T1]{fontenc}
\usepackage[greek, russian, main=english]{babel}

\usepackage{amssymb,amsfonts}
\usepackage{amsmath}

\usepackage{algorithm}
\usepackage{algorithmic}
\usepackage{caption}
\usepackage{graphicx}
\usepackage{textcomp}
\usepackage{xcolor}
\usepackage{adjustbox}
\usepackage{float}

\def\papertitle{Pairing Real-Time Piano Transcription \\with Symbol-level Tracking \\for Precise and Robust Score Following}

\author[1]{\mbox{\firstname{Silvan}\middlename{David}\lastname{Peter}}}
\author[1]{\mbox{\firstname{Patricia}\lastname{Hu}\originalname{胡紫漪}}}
\author[1, 2]{\mbox{\firstname{Gerhard}\lastname{Widmer}}}

\affil[1]{\institution{Institute of Computational Perception, Johannes Kepler University}\city{Linz}\country{Austria}\affiliationtype{University}}
\affil[2]{\institution{LIT AI Lab, Linz Institute of Technology}\city{Linz}\country{Austria}\affiliationtype{Music}}

\completesetup

\title{\papertitle}
\begin{document}

\capstartfalse
\maketitle
\capstarttrue

\begin{abstract}
Real-time music tracking systems follow a musical performance and at any time report the current position in a corresponding score.
Most existing methods approach this problem exclusively in the audio domain, typically using online time warping (OLTW) techniques on incoming audio and an audio representation of the score.
Audio OLTW techniques have seen incremental improvements both in features and model heuristics which reached a performance plateau in the past ten years.
We argue that converting and representing the performance in the symbolic domain -- thereby transforming music tracking into a symbolic task -- can be a more effective approach, even when the domain transformation is imperfect. 
Our music tracking system combines two real-time components: one handling audio-to-note transcription and the other a novel symbol-level tracker between transcribed input and score.
We compare the performance of this mixed audio-symbolic approach with its equivalent audio-only counterpart, and demonstrate that our method outperforms the latter in terms of both precision, i.e., absolute tracking error, and robustness, i.e., tracking success.
\end{abstract}

\section{Introduction}\label{sec:introduction}
The task of a real-time music tracking system (also known as a score follower) is to listen to a musical performance and recognize automatically, at any time, the current position in the corresponding musical score. 
Typically, the system possesses a representation of the score in advance, and maps an incoming real-time audio stream onto this representation to decode the current score position.
A system that achieves this task accurately and robustly promises to be useful in a wide range of applications, e.g., performance visualizations, live music subtitling, automatic score displays, or live synchronization between a musician and the score following system.

In a real-time music tracker, the representation of the two time series to be aligned -- score and live performance -- plays a crucial role.
Generally, there are two principal approaches: representing both time series in the audio domain and then aligning the two audio streams, or processing them in the symbolic domain, where music is represented as note events or MIDI data. 
The first approach is more versatile as both audio reference tracks (e.g., previous recordings of a piece)  as well as audio inputs (e.g., live recordings of an acoustic instrument) are more readily available than their symbolic representations such as MusicXML scores or MIDI performances.
Most methods to date have consequently followed the first approach, treating music tracking as an audio-to-audio alignment problem by synthesizing a MIDI version of the score into an audio file. 
Harmonic (energy distribution-based) or onset (energy dynamics-based) features are then used to align performance and score using online variants of the online time warping (OLTW) algorithm \cite{dixon2005line, arzt2012adaptive, arzt2008automatic, hu2003polyphonic,arzt2010simple,arzt2015real,Brazier2021lyrics,brazier2021improving,park2024phonetic,wei2018online,rodriguez2016tempo}.
Apart from these OLTW-based methods, especially earlier systems have relied on Hidden Markov Models (HMMs) and other Bayesian networks for alignment \cite{cont2009coupled, raphael2009current}.
The second approach, symbol-level tracking, is applied for high-precision scenarios such as automatic accompaniment\cite{cancino2023accompanion}.
While this domain has seen less research, symbol-level trackers tend to be more precise if successful, though they are more susceptible to errors from performance inconsistencies and reference mismatches\cite{cancino2023accompanion,peter2023online}

In this paper, 
we develop a novel symbol-level tracker based on online time warping (OLTW) that incorporates path-wise tempo estimation.
We evaluate this symbol-level tracker in 
a score following scenario where the real-time input consists of an audio recording and the score is provided in a symbolically encoded format.
Specifically, we demonstrate that transforming the incoming performance stream from the audio into the symbolic domain, and subsequently aligning it with its corresponding symbolic score, in real time, can surpass the performance of existing audio-based music tracking systems, even when the audio-to-symbolic conversion is imperfect.
We compare our method to its online audio-only counterpart and analyze how various tuning parameters, from feature selection to alignment algorithms, influence the performance.
Our experimental results highlight the potential of approaching real-time music tracking as a symbol-level task, supported by both improved, reliable symbol-level tracking and recent advances in real-time audio-to-note transcription.	
    
\section{Related Work}


Most existing systems have treated real-time music tracking as an online audio alignment problem, and developed their approaches from offline dynamic time warping (DTW).
In (offline) DTW, the two sequences to be aligned are completely known beforehand, and the goal is to find an optimal alignment between the two based on a local distance measure and an alignment cost matrix. 
Each entry in the matrix represents the cost of the optimal alignment up to that point, and 
the optimal warping path is found by recursively tracing back through the matrix using dynamic programming \cite{muller2021fundamentals}.

In the standard version of \textit{online} time warping (OLTW)\cite{dixon2005line}, only the score is known completely beforehand while the performance is streamed in real time.  
The alignment is now computed incrementally by expanding the window into the (row and/or column) direction with minimal normalized cost. 
This online variant achieves linear time and space complexity, as it updates only a fixed number of cells per step rather than the entire matrix. 
Variants of OLTW include enhancements such as the ‘backward-forward strategy’ \cite{arzt2008automatic} that revisits past decisions to improve the precision of the current score position estimate, the use of tempo models \cite{arzt2010simple} to adjust the score representation based on the tempo of the live performance, as well as the use of multiple reference performances\cite{arzt2015real,cancino2023accompanion}.

Most audio-based music tracking systems 
rely on carefully crafted audio features.
Widely used features include chroma-based harmonic features \cite{arzt2008automatic,arzt2010simple,raffel2016optimizing,muller2006efficient}, onset-based features\cite{ewert2009high}, or a combination of both, which balance chroma robustness with temporal accuracy \cite{arzt2012adaptive}. 
More recently, the focus has shifted toward learned features, leveraging deep neural networks \cite{bittner2017deep,arzt2018audio,dorfer2017learning, agrawal2021learning,park2024phonetic,Brazier2021lyrics,rodriguez2016tempo}.


Compared to audio-based music tracking, symbolic music tracking remains relatively under-explored\cite{peter2023online,cancino2023accompanion}.
In offline audio-to-score alignment, there is prior work that follows a similar approach to ours, that is, leveraging automatic music transcription (AMT) to predict note presence and onset timing, which are then used as note features for subsequent MIDI-based offline score alignment \cite{kwon2017audio, simonetta2021audio}. Moreover, some studies have explored a middle ground between audio and symbol-level alignment by aligning raw transcription features \cite{DBLP:conf/ismir/ZeitlerMM24}.

Recent advancements in AMT have been largely driven by the availability of large-scale, well-aligned audio-MIDI datasets and increased availability of computational resources \cite{hawthorne2019enabling}, enabling deep neural networks to achieve state-of-the-art performance \cite{kong2021highresolution, edwards2024data}.
However, as these models have grown in complexity and size, they have become increasingly resource-intensive, limiting their applicability in real-time scenarios. 
Most existing systems are therefore designed for offline use, with only a limited number addressing the task in real-time contexts \cite{jeong2020real,fernandez2023onsets,kusaka2024mobile}.

Online methods for AMT focus on reducing complexity through architectural choices such as single-branching \cite{jeong2020real} or substituting conventional components with more computationally lightweight alternatives \cite{kusaka2024mobile}, in addition to changing recurrent directions from bi- to uni-directional \cite{jeong2020real, kusaka2024mobile}, or removing recurrent layers entirely \cite{fernandez2023onsets}. 
	
\section{Proposed Method}
In this section, we introduce our score following method based on real-time piano transcription and symbol-level music tracking.
We start with the real-time transcription module which is chosen from the literature and adapted to our use case with minimal modifications. Most of this section then concerns our symbol-level tracking algorithm.

\subsection{Real-time AMT}
We choose to use \cite{kwon2020polyphonic} for the online transcription part of our method. 
\cite{kwon2020polyphonic} consists of a convolutional acoustic model that extracts local acoustic features, and a uni-directional LSTM that predicts notes per pitch and per frame auto-regressively. 
Compared to previous methods \cite{hawthorne2019enabling, kong2021highresolution}, the architecture in \cite{kwon2020polyphonic} is more lightweight as multiple note states (i.a., onset, off, on) are predicted in a multi-class framework rather than using separate model branches for each state. This model represents the current state of the art in online piano transcription, with competitive F1 scores at a latency of 174 ms.

\begin{algorithm}[t]
\caption{Pairwise Distance Metric combining Pitch and Time}\label{alg:metric}
\begin{algorithmic}[1]
\REQUIRE for score ($A$) and performance ($B$) each: current times $a^T_i, b^T_j$, previous times $a^T_{pi}, b^T_{pj}$, current pitch set $a^P_{i}$ and pitch $b^P_{j}$, previous tempo $t_p$. timing weight $c$, decay factor tempo $d$
\STATE pitch error $e^P \gets 0$ if $b^P_{j} \in a^P_{i}$ else $1$
\STATE current onset time estimate $\widehat{b^T_{j}} \gets b^T_{pj} + (a^T_i - a^T_{pi}) * t_p$ 
\STATE time error $e^T \gets |\widehat{b^T_{j}} - b^T_{j}|$
\STATE pairwise distance $pd \gets e^P + c * e^T$
\IF{$a^T_i \neq a^T_{pi}$}
    \STATE current tempo $t_c \gets (b^T_{j} + b^T_{pj}) / (a^T_i - a^T_{pi})$
\ELSE
    \STATE current tempo $t_c \gets t_p + b^T_{j} - b^T_{pj}$
\ENDIF
\STATE smoothed tempo $t_s \gets t_c * d + (1 - d) * t_p$
\RETURN $pd, t_s$
\end{algorithmic}
\end{algorithm}

\subsection{Symbol-level Tracking}
The symbol-level tracker builds on three separate strands of work: the core OLTW algorithm
used in audio tracking \cite{arzt2012adaptive,dixon2005line}, a pitch-based sequence representation introduced for offline symbol-level alignment \cite{peter2023online}, and tempo modeling \cite{cancino2023accompanion,rodriguez2016tempo,arzt2010simple}.
We combine these into a novel OLTW version on symbol sequences with a new pairwise pitch- and time-based distance metric.

Our tracker works on two sequences of symbolically encoded notes, each represented by integer MIDI pitches and floating-point onset times.
The performance ($B := [\dots,(b^P_j, b^T_j),\dots]$) is a sequence of pitch onset tuples, ($b^P$, $b^T$), representing MIDI notes, whether they are streamed in real-time, recorded, or transcribed.
The score sequence $A$, in contrast, encodes chords instead of individual pitches.
It is directly extracted from MIDI or a symbolically encoded score, grouping notes with coinciding onsets into pitch sets that represent all starting notes at each score onset.
The resulting score sequence ($A := [\dots,(a^P_i, a^T_i),\dots]$) consists of pitch set ($a^P$) and onset ($a^T$) tuples.

\begin{table*}[t]
 \centering
  \adjustbox{max width=\linewidth}{
\bgroup
\def\arraystretch{1.2}%
\begin{tabular}{|l|llllll|llllll|}
\hline
Window size & \multicolumn{6}{c|}{10} & \multicolumn{6}{c|}{20} \\
\hline
Weighting onset metric &\multicolumn{2}{c|}{0.5} & \multicolumn{2}{c|}{1.0} & \multicolumn{2}{c|}{2.0} &
\multicolumn{2}{c|}{0.5} & \multicolumn{2}{c|}{1.0} & \multicolumn{2}{c|}{2.0} \\
\hline
Weighting score direction & 1 & 2 & 1 & 2 & 1 & 2 & 1 & 2  & 1 & 2 & 1 & 2 \\
\hline
Robustness & 69.44 &   77.78 &   75.00 &   86.11 &   83.33 &   77.78 &   86.11 &   91.67 &   91.67 &   91.67 &   \textbf{91.67} &   88.89\\
\hline
\% abs error $\leq$ 0.5 s & 97.13 &   97.23 &   97.19 &   96.44 &   97.37 &   96.12 &   96.82 &   96.98 &   97.04 &   97.22 &   \textbf{97.26} &   96.95\\
\% abs error $\leq$ 0.25 s &   93.86 &   94.26 &   94.33 &   93.59 &   94.75 &   93.12 &   93.51 &   93.74 &   94.10 &   94.29 &   \textbf{94.72} &   94.07\\
\% abs error $\leq$ 0.1 s &   82.94 &   83.89 &   84.10 &   83.97 &   88.33 &   86.35 &   83.48 &   83.92 &   84.72 &   84.76 &   \textbf{85.93} &   84.66\\
\% abs error $\leq$ 0.05 s &  59.28 &   60.34 &   60.14 &   61.80 &   65.30 &   65.35 &   61.58 &   61.99 &   62.69 &   62.71 &   \textbf{63.70} &   62.86\\
\% abs error $\leq$ 0.025 s &   35.29 &   36.10 &   35.67 &   37.21 &   39.08 &   40.14 &   37.37 &   37.68 &   38.09 &   38.08 &   \textbf{38.64} &   38.59\\
\hline


\end{tabular}

\egroup}
 \caption{Robustness and precision results of parameter grid search of our symbolic OLTW tracker on \textit{transcribed} MIDI data, comparing effects of window size $w$, timing weighting $c$ in pairwise distance metric, and score direction penalization $dw_0$. 
 }
 \label{tab:tuning_sym_transcribed}
\end{table*}

\subsubsection{Pairwise distance metric}

Algorithm~\ref{alg:metric} describes the pairwise distance metric, which compares (performance and score) pitch and time (tempo) information.
Uppercase variables are used for sequences or matrices, lowercase ones for scalars and pitch sets. The pairwise distance metric computes a measure of closeness between an incoming performance tuple $(b^P_j, b^T_j)$ and a reference score tuple $(a^P_i, a^T_i)$, and consists of localized part based on pitch $e^P$, and context-dependent part $e^T$ based on both onset information and a tempo estimate $t_p$.

For the pitch error $e^P$ (line 1), we adapt the representation and metric in~\cite{peter2023online}, a binary distance between a performed pitch $b^P_{j} = p_j$ and a pitch set corresponding to a score onset $a^P_{i} = \{p_{i1},p_{i2},...\}$, which equals zero if $p_j \in a^P_{i}$, and one otherwise.
For the time error $e^T$, we first extrapolate the current performed onset time $\widehat{b^T_{j}}$ from the previous performance onset time, using the score inter-onset interval (i.e., the distance between previous and current score onset times) and the previous tempo estimate $t_p$ (line 2), and compare this estimate to the actually observed current performed note onset $b^T_j$ (line 3).
Intuitively, the time error $e^T$ measures how close the current performed onset time is to its expected position, based on the previous tracking position and tempo.
The total pairwise distance $pd$ is then computed as the sum of the pitch error $e^P$ and the time error $e^T$ weighted by factor $c$.
Besides the distance, Algorithm~\ref{alg:metric} also computes an updated smoothed tempo estimate $t_s$ for the current tracking position based an exponential moving average of previous tempo values with decay factor $d$ (lines 5 - 10).

\subsubsection{Adapted OLTW algorithm}

The distance metric usefully encapsulates both pitch and timing information. 
However, it relies on information about previous tracking positions as well as their tempo estimates, requiring modifications to the standard OLTW algorithm. Algorithm~\ref{alg:oltw} shows the central loop of our tracker that accommodates for this distance metric.

The underlying concept of OLTW is briefly outlined here, for a full description we refer to \cite{dixon2005line}. 
OLTW involves the iterative evaluation of the forward pass of a windowed dynamic time warping (DTW) algorithm.
Specifically, the accumulated cost between two windowed sequences is computed while tracking the path-wise alignment length. 
Normalization is applied to the costs along the outer border of the cost matrix by dividing them by their respective path lengths. 
The matching pair is then determined by selecting the sequence indices that correspond to the minimal normalized cost along the border. 
Following this, the window is extended in the direction of the match --- e.g., in the score direction if the last match is the final score note within the window --- and the process repeats. 
Core aspects of this algorithm, such as candidate selection, border selection, path normalization, and window initialization, are preserved in our symbol-level OLTW. 
For the sake of clarity, these steps are omitted from Algorithm~\ref{alg:oltw}.

The key novelties are twofold: 1) the introduction of directional weights $dw_0, dw_1, dw_2$ for the score, diagonal and performance directions, respectively, and 2) the incorporation of a tempo matrix $T$ (line 2) to track point (onset)-wise tempo estimates. 
The directional weights prioritize certain directions over others, thereby encouraging or penalizing specific paths. This technique is commonly employed in offline DTW to influence path selection.
In lines 6 - 8, the pairwise distances $pwc$ and tempo estimates $t$ corresponding to the three possible directions are computed based on the pairwise distance metric described above. 
Then, in lines 9 - 11, the distances are weighted by their directional weight and accumulated to directional costs $dc$.
The index $idx_{min}$ corresponding to the direction which minimizes the accumulated cost $dc$ is computed in line 12.
The cost at this index is stored in the windowed accumulated cost matrix $AC$ (line 13).
In line 14, the tempo estimate $t$ corresponding to the direction with the minimal accumulated cost is updated and stored in the tempo matrix $T$.
Rather than relying on an external tempo model or time-based heuristic, this algorithm implicitly tracks multiple alternative paths, each with its own continuously updated tempo estimate.
Finally, in line 15, the path length is updated and stored in its matrix $PL$.

The method involves several tuning parameters, including directional weights, initial tempo, tempo decay factor, window size, and timing weight, all of which impact its behavior, which are explored in further detail in the next section.

\begin{algorithm}[h!b]
\caption{Custom Online Time Warping Algorithm}\label{alg:oltw}
\begin{algorithmic}[1]
\REQUIRE reference sequence $A = \{a_1, a_2, \dots, a_n\}$, 
input queue $B = \{b_1, b_2, \dots, b_m\}$, 
local distance metric $m$, 
window size $w$,
directional weights $dw_0, dw_1, dw_2$.
\STATE $AC \gets AC_0$ 
\STATE $T \gets T_0$ 
\STATE $PL \gets 0$ 
\FOR{$b_k$ in $B.pop()$}    
    \FOR{$(i,j)$ in $idx_{border}$}
    \STATE $pwc_0, t_0 \gets m(a_i, b_j, a_{i-1}, b_{j}, T[i-1, j])$
    \STATE $pwc_1, t_1 \gets m(a_i, b_j, a_{i-1}, b_{j-1}, T[i-1, j-1])$
    \STATE $pwc_2, t_2 \gets m(a_i, b_j, a_{i}, b_{j-1}, T[i, j-1])$
    \STATE $dc_0 \gets dw_0 * pwc_0 + AC[i-1, j]$
    \STATE $dc_1 \gets dw_1 * pwc_1 + AC[i-1, j-1]$
    \STATE $dc_2 \gets dw_2 * pwc_2 + AC[i, j-1]$
    \STATE $idx_{min} \gets $argmin$(dC_0,dC_1,dC_2)$
    \STATE $AC[i,j] \gets [dc_0,dc_1,dc_2][idx_{min}]$
    \STATE $T[i,j] \gets [t_0,t_1,t_2][idx_{min}]$
    \STATE $PL[i,j] \gets 1 + [PL[i-1,j],PL[i-1,j-1]PL[i,j-1]][idx_{min}]$
    \ENDFOR
\ENDFOR
\end{algorithmic}
\end{algorithm}
	
\section{Experiments}

\subsection{Experimental setup}

We evaluate our real-time music tracking method on the Batik-plays-Mozart corpus \cite{hu2023batik}, a dataset of precise, note-aligned MIDI piano performances and MusicXML scores.
We complete this public symbolic data with the carefully aligned original audio recordings of the performances.
To enable audio-to-audio score following we also add synthesized versions of the scores, using Fluidsynth\footnote{\url{https://www.fluidsynth.org/}} at 44.1 kHz with a standard tempo of 120 quarter notes per minute.

In a first experiment, we examine the impact of tuning parameters on our method. 
Specifically, we evaluate four window sizes, $w \in {10, 20, 40, 80}$, timing weights, $c \in {0.5, 1.0, 2.0}$, and directional weights, which penalize the score direction by a factor of $dw_0 \in \{1, 2\}$.
The other directions which are weighted at one ($dw_1 = 1$ and $dw_2 = 1$), a step in score direction is either twice as costly ($dw_0 =2 $) or equal ($dw_0 = 1$).
We set the tempo decay to $d = 0.1$ and the initial tempo to 0.5 seconds per quarter note. To assess the influence of different parameter settings, we evaluate our method on both \textit{transcribed} and \textit{recorded} MIDI performance data. The latter serves as an approximation of the upper performance bound of our online symbolic tracker, assuming an ideal real-time transcription that accurately detects all onsets.

In a second experiment, we compare our method to its real-time audio-only counterpart. Our audio-only tracking is based on a reference OLTW implementation from the matchmaker Python package \cite{matchmaker_lbd}, with modifications and fine-tuned parameters to ensure competitiveness with state-of-the-art audio tracking.

First, we modify the OLTW method to support weighted directions. Second, we evaluate a range of audio features: normalized semitone spectra (NS), normalized chroma (NC), locally normalized semitone onset (LNSO), and locally normalized chroma onset (LNCO). For NS and NC, frame-wise distances are computed using the Euclidean distance, whereas for LNSO and LNCO, a normalized and weighted L1 metric is employed as a feature-specific distance measure \cite{arzt2012adaptive}.

All features are computed from a constant-Q transform with 88 semitone bins, using a hop size of 1024 on 44.1 kHz audio with Hann windows. The OLTW window size is set to 500 frames (11.6 seconds), with an average of 89 performance onsets per window in our dataset.
Since OLTW performs best when tempos are similar \cite{arzt2010simple}, we stretch the score by the nearest integer ratio of score to performance length. Additionally, we experiment with two different directional weight settings, where the diagonal is encouraged either weakly ($dw_1 = 0.9$) or strongly ($dw_1 = 0.5$) while the other directions are kept at one ($dw_0 = 1$ and $dw_2 = 1$).

For evaluation, we employ two metrics: robustness and precision.
We define \textit{robustness} as the proportion of performances in the dataset that the score follower successfully tracks until the end without losing alignment. Detecting when a tracker is lost is nontrivial; therefore, we apply a heuristic criterion, considering a tracker lost if its absolute error exceeds 10 seconds at any point.
\textit{Precision} is quantified as the absolute error between predicted and actual performance times, computed for each score onset. To present the distribution of absolute errors, we report five percentages: the proportions of errors below 25 ms, 50 ms, 100 ms, 250 ms, and 500 ms, respectively.
Precision values are computed only for robustly tracked performances, as the large errors from lost trackers are uninformative and would therefore distort the analysis.

\subsection{Results}

Table~\ref{tab:tuning_sym_transcribed} shows an excerpt of the results of the parameter combinations addressed in the first experiment. 
Window sizes $w>20$ are omitted (from the table) as they neither improve precision nor robustness.
Trackers with $w=10$ are significantly less robust while achieving precision comparable to those with $w = 20$.
For recorded MIDI data (see Table ~\ref{tab:tuning_sym_original}), a window size of 40 results in increased robustness, while precision values remain largely unchanged.

We found $w = 20$, $c = 2.0$, and $dw_0 = 1$ to be the best parameter configuration as it maximizes precision under the constraint of maintaining strong robustness.
While the effect of the timing weight $c$ (see Algorithm~\ref{alg:metric} for details) is less clear, higher time weighting seems to generally improve the performance.
The interaction between timing and score directional weight is generally positive, with increased penalization improving tracking in most configurations. However, in the selected best configuration, this interaction unexpectedly results in a slightly negative effect.

\begin{table}[t]
 \centering
 \adjustbox{max width=\columnwidth}{
 \bgroup
\def\arraystretch{1.2}%
\begin{tabular}{|l|llll|}
\hline
Window size & \multicolumn{2}{c|}{20} & \multicolumn{2}{c|}{40} \\
\hline
Weighting onset metric  & \multicolumn{4}{c|}{2.0}  \\
\hline
Weighting score direction & 1 & 2 & 1 & 2 \\
\hline
Robustness &  88.89 &   91.67 &   94.44 &   \textbf{94.44}\\
\hline
\% abs error $\leq$ 0.5 s  &   98.55 &   98.67 & 98.12 &   \textbf{98.46} \\
\% abs error $\leq$ 0.25 s &    97.70 &   97.79 &  97.15 &   \textbf{97.52}\\
\% abs error $\leq$ 0.1 s &     96.19 &   96.36 & 95.63 &   \textbf{96.09}\\
\% abs error $\leq$ 0.05 s &     94.71 &   94.76 & 94.16 &   \textbf{94.54}\\
\% abs error $\leq$ 0.025 s &  93.50 &   93.53 &  92.95 &   \textbf{93.32}\\
\hline
\end{tabular}
\egroup}
 \caption{Grid search results excerpt for OLTW tracker on \textit{recorded} MIDI data. Parameter configurations identical as those tested in Table \ref{tab:tuning_sym_transcribed}.
 }
 \label{tab:tuning_sym_original}
\end{table}

Table~\ref{tab:baseline_oltw} presents an excerpt from the grid search results of an audio-based OLTW baseline. We report results for directional weights $dw_1 = 0.5$, which strongly favor the diagonal direction, as these yielded the best performance in our experiments. Overall, the values range in the same order of a highly tuned OLTW model \cite{arzt2012adaptive}.

Comparing the features, harmonic features (NC and NS) demonstrate greater robustness, while onset-based features (LNCO and LNSO) show higher precision, particularly at low error thresholds. 
This trade-off is in line with our understanding of the inherent characteristics of these feature types.
All audio trackers show notably lower robustness and precision than our proposed method. 
The only exception is the precision of onset-based features at low error thresholds, which slightly outperforms our best configuration in terms of precision. However, this improvement in precision comes at the cost of a considerably lower robustness, i.e., tracking success rate.

It is important to note that the precision of our proposed method is bounded by the resolution of the transcription model, which operates at a 32 ms hop size.
The audio OLTW model, on the other hand, has a hop size of approximately 23 ms. 
Even if these trackers identify the correct matching frames for each note, the absolute error might still be as large as 16 ms or 23 ms, respectively\footnote{
The ground truth alignment stems from symbolically encoded scores and performances.
If a MIDI note is correctly transcribed, i.e., the audio frame containing its onset is correctly identified by the transcription model, the maximal error between the frame center and the ground truth MIDI note onset is half the hop size, or 16 ms.
For the audio matching case, the maximal symbol-to-audio frame error of half a hop size may occur twice, once for the score and once for the performance. 
As a result, the maximum total error between ground truth and correctly matched frame is one full hop size, or 23 ms.
}. 
Thus, while it is possible, it is very difficult to achieve the highest precision quantile of 25 ms.

However, we still include the absolute error at this precision, not least to evaluate the performance of symbolic tracking approach in isolation. 
The tracking results on recorded MIDI in Table ~\ref{tab:tuning_sym_original} represent an upper bound for the symbolic tracker, assuming perfect transcription conditions. 
Notably, the precision at low error thresholds is significantly higher than what is typically achievable with audio trackers.
High quantiles clearing these low error thresholds are particularly important for reactive, musical tasks like accompaniment generation.

Although our method is precise and robust, it's real-time application is  complicated by the (fixed) latency of the transcription model of 174 ms.
In practice, this can be addressed by predicting current onsets 174 ms into the future. This extrapolation step typically results in a slight reduction in alignment precision, yet it is not uncommon in real-time tracking. 
The Disklavier series of computer controlled grand pianos has a fixed mechanical delay in the order of 200 ms which requires prediction of accompaniment in the future\cite{cancino2023accompanion}.
For less time-critical contexts such as subtitling or page turning, the current configurations are sufficiently fast without the need for adaptation or extrapolation.

\begin{table}[t]
\centering

\adjustbox{max width=\columnwidth}{\bgroup
\def\arraystretch{1.2}%
\begin{tabular}{|l|llll|}
\hline
Weighting diagonal direction & \multicolumn{4}{c|}{0.5}  \\
\hline
Feature & NS & NC & LNSO & LNCO \\
\hline
Robustness & 88.89 &   88.89 &   58.33 &   72.22 \\
\hline
\% abs error $\leq$ 0.5 s &  94.98 &   95.07 &   93.19 &   87.46 \\
\% abs error $\leq$ 0.25 s &  89.04 &   89.37 &   87.59 &   81.83 \\
\% abs error $\leq$ 0.1 s &   71.44 &   73.02 &   77.42 &   74.38 \\
\% abs error $\leq$ 0.05 s &   49.30 &   52.11 &   63.30 &   60.70 \\
\% abs error $\leq$ 0.025 s & 28.75 &   32.02 &   45.33 &   44.91 \\
\hline
\end{tabular}

\egroup}
\caption{Robustness and precision of the baseline audio OLTW music tracker with diagonal weight ($dw_1=0.5$) using different chroma-based (NS and NC) and onset-based (LNSO and LNCO) features.}
\label{tab:baseline_oltw}
\end{table}

\section{Conclusion}

In this work, we introduced a score following method that operates in the symbolic domain, tackling the problem using real-time transcription and a novel symbol-level online tracker. Our system tracks complex piano music with high robustness and precision, and even does so in the absence of perfect transcription.

Our method is independent of the specific music transcription algorithm used, so our tracking framework will automatically benefit from ongoing work (e.g., \cite{kusaka2024mobile}) or future improvements in the field of real-time music transcription:
In particular, higher transcription fidelity and lower latency would significantly enhance the performance of our model, making it well-suited for real-time accompaniment generation.
Considering that our symbolic tracker achieves high robustness and precision on recorded MIDI data -- representing an upper bound of this method -- we argue that this transcribe-and-track approach is a promising direction for future research.

\begin{acknowledgments}
This research acknowledges support by the European Research Council (ERC), under the European Union's Horizon 2020 research and innovation programme, grant agreement No.~101019375 \textit{Whither Music?}.
We thank Jiyun Park and Carlos Cancino-Chac\'on for their valuable insights and constructive discussion.
\end{acknowledgments} 

\bibliography{refs}
	
\end{document}